\documentclass[aip,jcp,reprint,amsmath,amssymb]{revtex4-1}

\usepackage{graphicx}
\usepackage{dcolumn}
\usepackage{bm}
\usepackage[dvips]{color}
\usepackage{subfigure}

\newcommand{\rvec}{\textit{\textbf{r}} }
\newcommand{\avec}{\textit{\textbf{a}} }
\newcommand{\bvec}{\textit{\textbf{b}} }
\newcommand{\xvec}{\textit{\textbf{x}} }
\newcommand{\nvec}{\textit{\textbf{n}} }
\newcommand{\ivec}{\textit{\textbf{i}} }
\newcommand{\jvec}{\textit{\textbf{j}} }
\newcommand{\Rvec}{\textit{\textbf{R}} }
\newcommand{\kvec}{\textit{\textbf{k}} }
\newcommand{\Ymat}{\textbf{Y} }
\newcommand{\Amat}{\textbf{A} }
\newcommand{\Bmat}{\textbf{B} }

\begin{document}

\title{Vesicle deformations by clusters of transmembrane proteins}

\author{Amir Houshang Bahrami}
\author{Mir Abbas Jalali}
\email{mjalali@sharif.edu}
\affiliation{Department of Mechanical Engineering,
             Computational Mechanics Laboratory \\
             Sharif University of Technology, Azadi Avenue, Tehran, Iran}

\date{\today}

\begin{abstract}
We carry out a coarse-grained molecular dynamics simulation of phospholipid 
vesicles with transmembrane proteins. We measure the mean and Gaussian curvatures 
of our protein-embedded vesicles and quantitatively show how protein clusters 
change the shapes of their host vesicles. The effects of depletion force and 
vesiculation on protein clustering are also investigated. By increasing the protein 
concentration, clusters are fragmented to smaller bundles, which are then 
redistributed to form more symmetric structures corresponding to lower bending 
energies. Big clusters and highly aspherical vesicles cannot be formed when 
the fraction of protein to lipid molecules is large.  
\end{abstract}

\maketitle

\section{Introduction}

Biological membranes are found in various complex 
shapes \cite{Lip95,Sei97} which are closely related to membrane 
functions such as biconcave shape of erythrocytes or corkscrew 
shape of spirochetes. Shape variety depends mainly on protein 
concentration, operation of external forces on membranes in 
cellular environments \cite{Zim06}, membrane movement, fusion 
and budding processes, variations in lipid composition, and 
vesicle trafficking \cite{Mc05}. Membrane local curvature is 
representative for its shape, and proteins are believed to play 
an important role in membrane conformation. Proteins behave as 
both generating \cite{Zim06,Mc05} and sensing \cite{Vog06} 
elements for the membrane curvature \cite{Mc05,Phi09}. While 
protein aggregation induces shape transformation, membrane 
curvature may also generate feedback on protein aggregation 
and yield attractive \cite{Phi09,Rey07} or repulsive 
\cite{Phi09,Kim98} curvature-mediated interactions between 
them. It has also been shown that bending rigidity and membrane 
thickness affect protein functioning \cite{And07,Jen04}. 

Proteins affect membrane curvature through various ways 
such as scaffold and local curvature mechanisms, and by their
integration (as transmembrane proteins) with the membrane 
\cite{Zim06,Mc05,Phi09,Koz10}. However, the experimental 
measurements of variations in the membrane curvature are not 
easy. The role of proteins on the membrane deformations has 
been studied using continuum elastic modeling \cite{Phi09,Kim98,Wei98}, 
particle-based \cite{Cha08} and mesoscopic \cite{Ven05} simulations, 
and hybrid elastic-discrete particle models \cite{Naj09}. In all 
these studies, lipid bilayers represent a liquid environment with 
freely diffusing lipid chains that dissolve topological deformations. 
The aggregation of somewhat rigid proteins remarkably reduces the 
diffusion of lipid chains and causes shape variations. 

In this study, we use a coarse-grained model \cite{Ven06} and 
generalize the method of Markvoort et al. \cite{Mar05} to generate 
phospholipid vesicles from initially rectangular and flat bilayers 
with different concentrations of transmembrane proteins. We discuss 
our simulation method in section \ref{sec:simulation-procedure} and 
model the vesicle surface using spherical harmonics. In section 
\ref{sec:curvatures}, we present a method for computing the mean 
and Gaussian curvatures at the locations of hydrophilic heads of 
lipid chains and proteins. The local curvature information together 
with the sizes of protein clusters help us to investigate the 
deformations of host vesicles in section \ref{sec:examples}. 
We also study the size distribution, formation and fragmentation 
of protein clusters. We summarize our fundamental results in 
section \ref{sec:conclusions}.

\section{Model Description}
\label{sec:simulation-procedure}

Mesoscopic models have been widely used to study the physics of 
membranes \cite{Ven06}. Lipids can spontaneously aggregate and form 
various membranes \cite{Goe98,Goe99}. When lipid chains are assembled 
in the form of a closed 3D surface and trap water molecules, a vesicle 
is generated. Entropy is the main driving mechanism of this process
\cite{Mar05}. To construct vesicles, we insert initially flat rectangular 
lipid bilayers (which may contain proteins) inside a box of water molecules. 
Such a configuration is unstable because the tails 
of boundary lipids are repelled by water molecules and the bilayer 
is compressed by in-plane forces. Consequently, the bilayer buckles 
and closes itself to acquire a minimum potential energy state. Vesicles 
formed through this bilayer $\rightarrow$ vesicle transition process, 
with the progenitor bilayer being surrounded by solvent particles 
(without touching simulation boundaries), have a more relaxed pressure 
distribution. Furthermore, using bilayer $\rightarrow$ vesicle transition 
process we obtain vesicles of different sizes in a more controllable 
process.

\begin{figure}
\begin{center} \mbox{\subfigure(a)
{\includegraphics[angle=-90,width=0.10\textwidth]{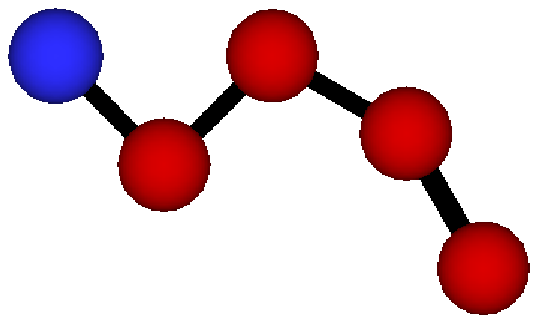}}\quad
\subfigure(b)
{\includegraphics[angle=-90,width=0.13\textwidth]{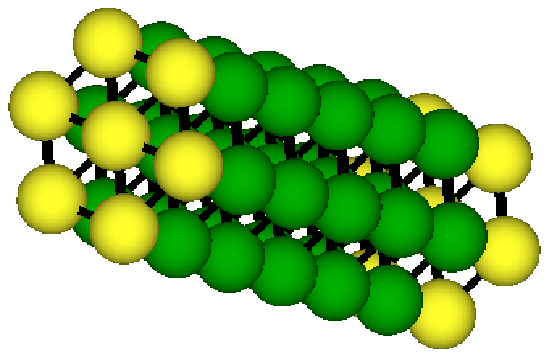}}
}
\caption{(a) A flexible lipid chain with one hydrophilic head (blue) 
and four hydrophobic tail particles (red). (b) Protein molecules consisting 
of seven strands. Each strand has two hydrophilic ends (yellow particles) 
and five hydrophobic particles in the middle (green). Particles of neighboring 
strands are connected by elastic rods (black lines).}
\label{fig:fig1}
\end{center}
\end{figure}

Our simulation box has dimensions of $L_x\times L_y\times L_z$ with 
the $x$-axis being normal to the initial bilayer mid plane. We use periodic 
boundary conditions, and choose a sufficiently large box so that the bilayer 
does not reach to the boundaries before vesiculation. The number of protein 
and lipid molecules are denoted by $N_p$ and $N_l$, respectively. The particles 
that constitute the elements of our setups either are water particles (type 1), 
or have hydrophobic (type 2) and hydrophilic (type 3) natures. 
We follow reference [19], and model each lipid chain by one hydrophilic head 
and four hydrophobic tail particles (Fig. \ref{fig:fig1}{\em a}).
All particles interact through Lennard-Jones (LJ) potential \cite{Allen87} 
with a cut off radius $r_c = 2.5 \sigma$. 

We use the integer subscripts $i$ and $j$ for particle types, and 
index a particle in a molecule by the subscript $s$. For the pairs 
$(i,j)=(1,2)$ and (2,3) the interaction potential is 
$V_{ij} = 4\epsilon_{ij}(\sigma_{ij}/r)^{9}$, and for other pairs we 
use $U_{ij} = 4\epsilon_{ij}[(\sigma_{ij}/r)^{12}-(\sigma_{ij}/r)^{6}]$
where $\sigma_{ij},\epsilon_{ij}$=1 $(i,j=1,2,3)$ \cite{Goe98}.
A harmonic bond potential of the form $U^{\rm bond}_{s,s+1} = K_{b}(\left|r_{s,s+1}\right|-\sigma)^{2}$
is applied between neighboring particles inside a lipid chain, and 
we have set $K_{b}=5000 \epsilon_{s,s+1} \sigma^{-2}$ where $\epsilon_{s,s+1}=1$.
With these assumptions, less than 10\% of bond lengthes fluctuate more 
than 2\% around $\sigma$ [19]. In some case studies, to include bending 
rigidity in our lipid and protein chains, we use the potential
\begin{eqnarray}
U^{\rm bend}_{s-1,s,s+1} &=& 
K_{\rm bend} \left (\! 1 \!-\! \frac{\rvec_{s-1,s}.\rvec_{s,s+1}}
                 {\left|\rvec_{s-1,s}\right|\left|\rvec_{s,s+1}\right|} \right ), \\
&=& K_{\rm bend}(1-\cos\phi_s),
\end{eqnarray}
where $K_{\rm bend}$ is the bending coefficient of non-flexible chains 
\cite{Goe98}. $\phi_s$ defines the bending angle between adjacent bonds 
in a single chain. Protein molecules (Fig. \ref{fig:fig1}{\em b}) are 
composed of seven strands in a hexagonal arrangement \cite{Ven05}. 
A particle in each strand interacts with all of its neighbors, within 
the same protein, through the potentials $U^{\rm bond}_{s,s+1}$ and 
$U^{\rm bend}_{s-1,s,s+1}$ defined above. In this way, lipid-protein 
interactions are the same as lipid-lipid interactions.

In this study, we work with simple proteins whose lengths are adjusted 
in a way that the hydrophobic mismatch effect is minimum, and such that 
a single protein does not affect the membrane conformation considerably. 
Shape variations that we report, are thus caused {\it only} by clustering.
Both flexible and rigid chains can be used in lipid and protein molecules. 
The experiments of section \ref{sec:examples} show that adding bending 
rigidity does not induce remarkable qualitative or quantitative changes 
in the protein tilting angle or in the aggregation products. In fact, the 
strong harmonic bond potentials assumed between neighboring strands 
(in a protein molecule) provide enough bending rigidity for our relatively 
short protein molecules.

We implement equilibrium molecular dynamics simulation of an NVT ensemble 
\cite{Allen87} with velocity Verlet algorithm for integration in the time 
domain. We use the integration time step $\delta t = 0.005 t_{0}$ with 
\begin{eqnarray}
t_{0} = \sqrt{\frac{m \sigma^2}{48 k_{\rm B} T}} = \frac{1}{\sqrt{64.8}},
~ k_{\rm B}=1,~ T=1.35,
\end{eqnarray}
and set the number density of particles\cite{Goe98} to $\rho = 2/3$.
All Particles have equal masses of $m=1$, and all lengths are scaled
by $\sigma$. The parameters used in our models produce correct physical 
properties of bilayers, including diffusion coefficients, density profiles 
and mechanical properties like surface tension and stress distribution
\cite{Goe98,Mar05}. After vesicle formation, the position vectors of 
particles are measured with respect to the vesicle center, and the 
lipid or protein heads exposed to water molecules outside the vesicle 
are tagged as surface particles. We assign an integer number $n$ to 
each surface particle, and denote the total number of surface particles 
by $N$. It is remarked that there is not a meaningful correlation 
between the physical location of each particle and its number $n$. 
The identifier $n$ is used only for statistical purposes. Furthermore, 
a single number $n$ is assigned to each surface particle, i.e., there 
are, respectively, one and seven particle identifiers corresponding 
to each lipid chain and protein molecule.

\section{Local mean and Gaussian curvatures}
\label{sec:curvatures}

To measure the local curvature of vesicles, with and without proteins, 
we express the radial distance of surface particles from the vesicle 
center in terms of spherical harmonics \cite{Arf85} as
\begin{eqnarray}
r(\theta,\phi)=\sum_{l=0}^{l_{\rm max}}\sum_{m=0}^{l} \left [ a^m_l A^m_l(\phi,\theta)
+ b^m_l B^m_l(\phi,\theta) \right ],
\label{eq:position-vector}
\end{eqnarray}
where 
\begin{eqnarray}
A^m_l &=& {\rm Re} \left [ Y^m_l \right ],~
B^m_l={\rm Im} \left [ Y^m_l \right ], \\
Y^m_l &=& (-1)^m e^{{\rm i}m\phi}
\sqrt{\frac{(2l+1)(l-m)!}{4\pi(l+m)!}} P^m_l(\cos\theta).
\end{eqnarray}
Here $P^m_l$ are associated Legendre functions and ${\rm i}=\sqrt{-1}$.
We find the coordinates of surface particles $(r_n,\phi_n,\theta_n)$,
define
\begin{eqnarray}
\Rvec = \left [
\begin{array}{cccc}
r_{1}(\phi_{1},\theta_1) ~ & r_{2}(\phi_{2},\theta_2) ~ & \ldots ~ &
                 r_{N}(\phi_N,\theta_{N} )
\end{array}                 
    \right ]^{\rm T},            
\label{eq:r-vector}
\end{eqnarray}
and collect all constants coefficients $a^m_l$ and $b^m_l$ in single 
column vectors $\avec$ and $\bvec$, respectively. The superscript 
T means transpose. We also define the matrices $\Amat$ and $\Bmat$ 
whose elements are $A^m_l(\phi_n,\theta_n)$ and $B^m_l(\phi_n,\theta_n)$, 
respectively. The discrete form of equation (\ref{eq:position-vector}) 
thus becomes 
\begin{equation}
\Rvec=\Ymat \cdot \xvec,~~
\Ymat = \left [ 
\begin{array}{cc}
\Amat & \Bmat
\end{array}
\right ], ~~ \xvec=\left \{
\begin{array}{c}
\avec
 \\ 
\bvec
\end{array}
\right \},
\label{eq:reduced-matrix-equation}
\end{equation}
which in practice, has more equations than unknowns. 

We calculate $\xvec$ using the singular value decomposition of $\Ymat$, 
and obtain the position of any surface particle from (\ref{eq:position-vector}) 
and 
\begin{eqnarray}
\rvec = r\left [ \sin (\theta)\cos (\phi) \hat \ivec +
                 \sin (\theta)\sin (\phi) \hat \jvec +
                 \cos (\theta) \hat \kvec \right ],
                 \label{eq:define-r-vs-phi-theta}
\end{eqnarray}
where $(\hat \ivec,\hat \jvec,\hat \kvec)$ are the unit vectors in Cartesian 
coordinates. Let us define $\rvec_{\phi}$ and $\rvec_{\theta}$ as the first order, 
and $\rvec_{\phi \phi}$, $\rvec_{\phi \theta}$, and $\rvec_{\theta \theta}$ 
as the second order partial derivatives of $\rvec$ in (\ref{eq:define-r-vs-phi-theta}) 
with respect to $\phi$ and $\theta$. The coefficients of the first fundamental 
form of the surface are thus determined as
\begin{eqnarray}
E=\rvec_{\phi} \cdot   \rvec_{\phi},~~
F=\rvec_{\phi} \cdot   \rvec_{\theta},~~
G=\rvec_{\theta} \cdot \rvec_{\theta}.
\end{eqnarray}
Defining the unit vector normal to the vesicle surface as 
$\hat \nvec = (\rvec_{\phi} \times \rvec_{\theta})/|\rvec_{\phi}\times \rvec_{\theta}|$,
one finds the coefficients of the second fundamental form:
\begin{eqnarray}
L=\rvec_{\phi\phi} \cdot \hat \nvec,~~
M=\rvec_{\phi\theta} \cdot \hat \nvec,~~
N=\rvec_{\theta\theta} \cdot \hat \nvec.
\end{eqnarray}
The mean curvature $H$ and the Gaussian curvature $K$ at the location of 
each surface particle can thus be computed using \cite{Saf94}
\begin{eqnarray}
H=\frac{LN-M^2}{EG-F^2},~~K=\frac{EN-2FM+GL}{2(EG-F^2)}.
\label{eq:K-and-H}
\end{eqnarray}
The principal curvatures ($C_{1},C_{2}$) are related to $H$ and $K$
through $H=(C_{1}+C_{2})/2$ and $K=C_{1} C_{2}$ \cite{Saf94}. 
In our numerical experiments we have truncated the series 
(\ref{eq:position-vector}) at $l_{\rm max}=5$. Including $l_{\rm max}>5$ 
terms had $\approx 3\%$ improvement in fractional errors. 
To perform a global shape classification of vesicles, we compute 
the average curvatures
\begin{eqnarray}
\left \{ 
\begin{array}{l}
\bar H \\ 
\bar K
\end{array}
\right \}
&=& \frac{1}{N} \sum_{n=1}^{N}
\left \{ 
\begin{array}{l}
H(n) \\ 
K(n)
\end{array} 
\right \},
\label{eq:averaged-curvatures}
\end{eqnarray}
and their corresponding standard deviations $\tilde H$ and 
$\tilde K$ for particles living on the surface of model vesicles.

\begin{figure*}
\begin{center} 
{\includegraphics[angle=-90,width=0.2\textwidth]{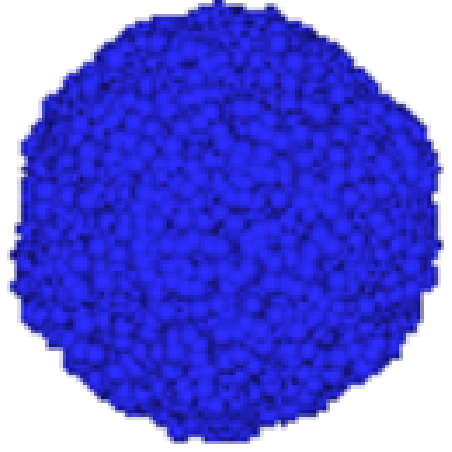}}
{\includegraphics[angle=-90,width=0.2\textwidth]{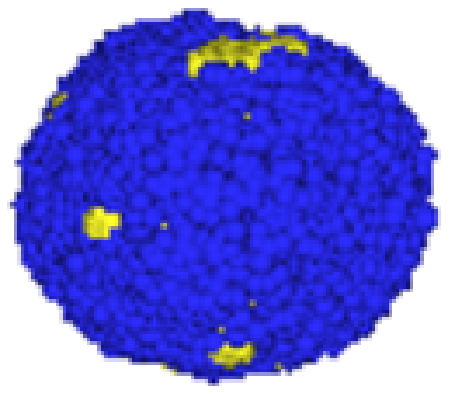}}
{\includegraphics[angle=-90,width=0.2\textwidth]{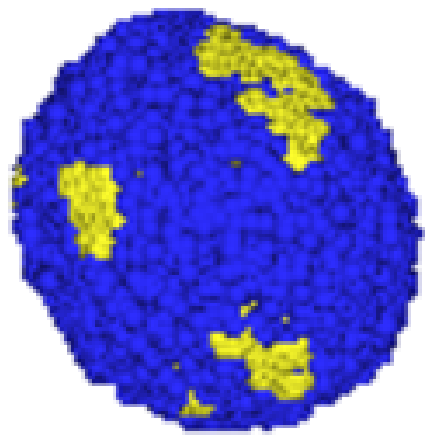}}
{\includegraphics[angle=-90,width=0.2\textwidth]{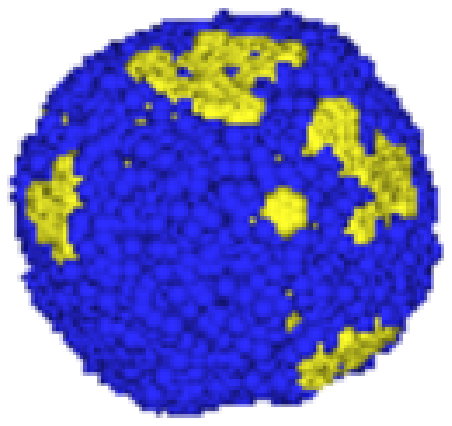}}\\
{\includegraphics[angle=-90,width=0.2\textwidth]{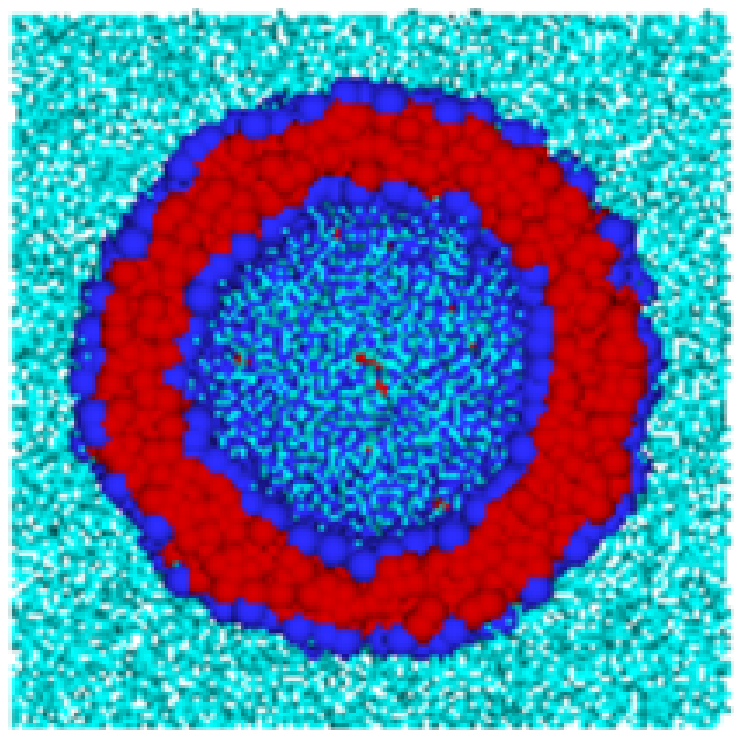}}
{\includegraphics[angle=-90,width=0.2\textwidth]{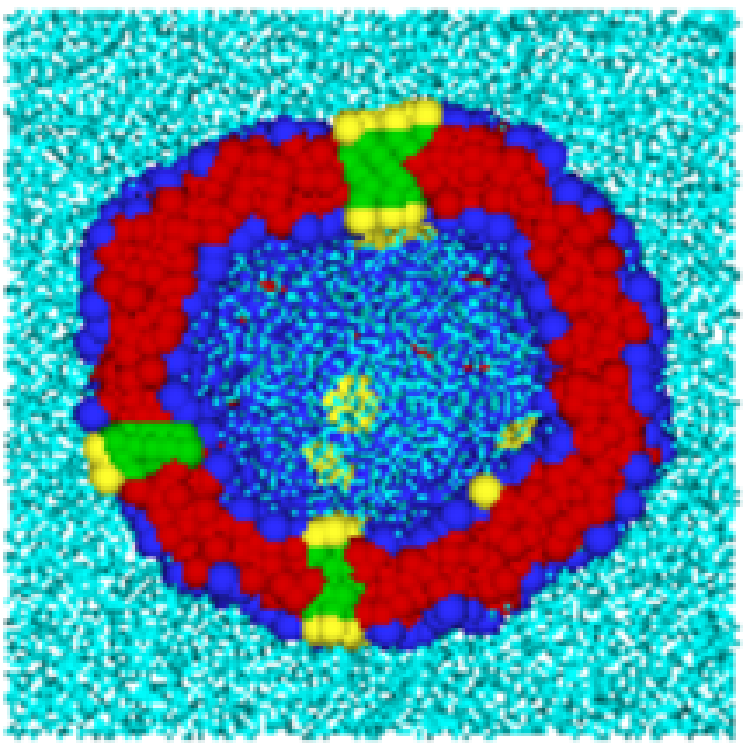}}
{\includegraphics[angle=-90,width=0.2\textwidth]{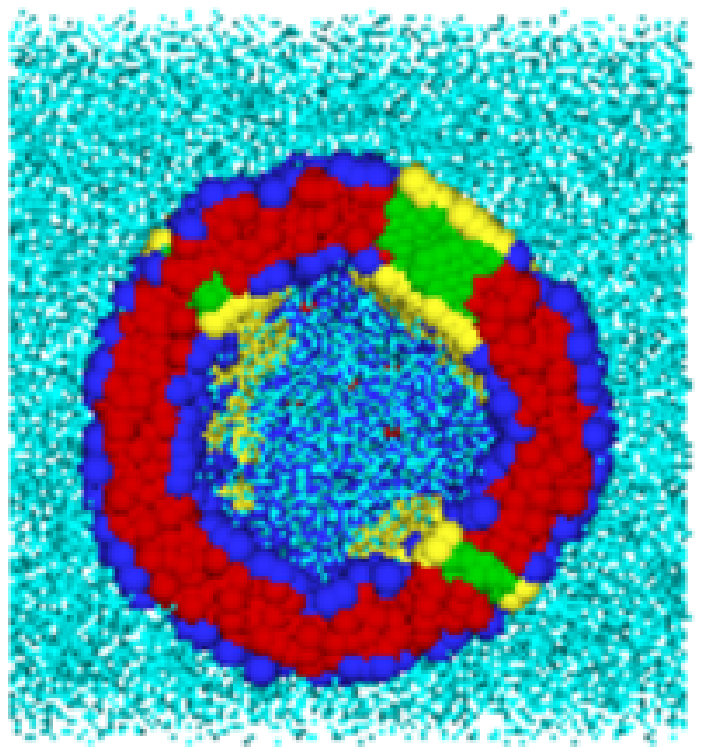}}
{\includegraphics[angle=-90,width=0.2\textwidth]{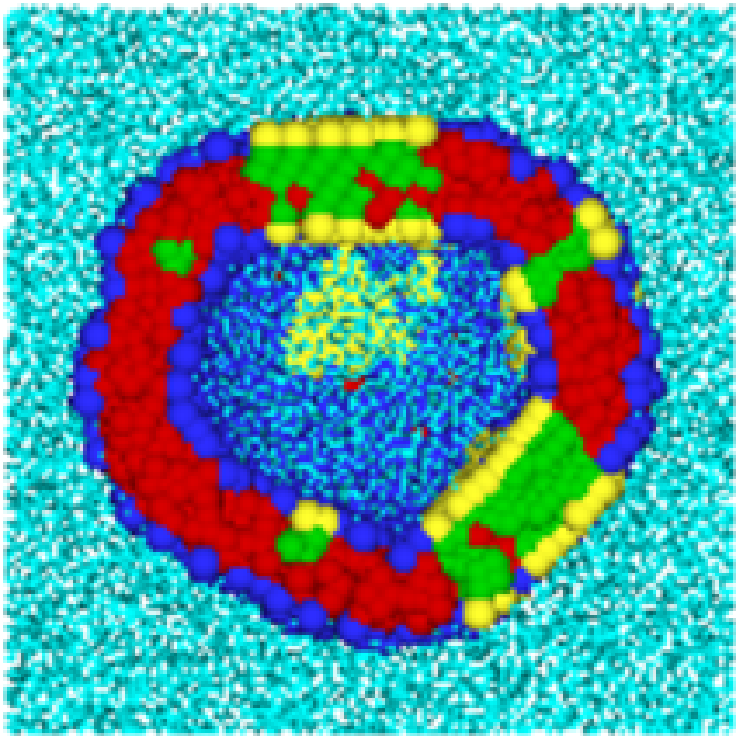}}
\caption{Three dimensional views (top row) and the most informative
cross sections (bottom row) of our simulated vesicles without 
(model $V_1$) and with transmembrane proteins. From left to right: 
models $V_1$, $V_2$, $V_3$ and $V_4$.}
\label{fig:fig2}
\end{center}
\end{figure*}

\section{Simulation Results}
\label{sec:examples}

We first consider the case of lipid and protein chains with no 
bending rigidity. Our first model, which is called $V_1$, is a vesicle 
without proteins and it is composed of $N_l=2300$ similar lipid chains. 
The initial flat bilayer is bent inside the water (solvent) particles 
and gradually forms an almost spherical vesicle shown in the left 
panels of Fig. \ref{fig:fig2}. To measure the sphericity of this 
vesicle quantitatively, we use a spherical harmonics expansion with 
$l_{\rm max}=5$, and fit a 3D surface to the surface particles.  
The mean local curvature $H(n)$ is computed from (\ref{eq:K-and-H}) and 
used in (\ref{eq:averaged-curvatures}) to find $\bar H=0.0564$ and 
$\tilde H=0.0068$. The small value of $\tilde H$ compared to $\bar H$ 
confirms the spherical nature of vesicle $V_1$.

\subsection{Protein-embedded vesicles}
\label{subsec:protein-embedded-vesicles}

We replace some lipid chains of our initially flat bilayer by protein 
molecules while keeping the area of generating bilayer almost constant. 
We randomly distribute proteins in the bilayer sheet, but observe that 
they form small clusters after the bending of membrane and during vesicle 
formation. The vesiculation process takes from $\approx 5\times 10^{5} \delta t$ 
for model $V_{1}$ to $\approx 2\times 10^{6}\delta t$ for protein-embedded 
vesicles. That is because proteins (or clusters of proteins) resist against 
buckling by decreasing the fluidity of the progenitor membrane. After vesicle 
formation, we have waited for about $5\times 10^{5}$ time steps to ensure 
that vesicles have reached to an equilibrium condition so that the number 
and area of protein clusters remain constant.

\begin{figure}
\begin{center}
{\includegraphics[width=0.2\textwidth]{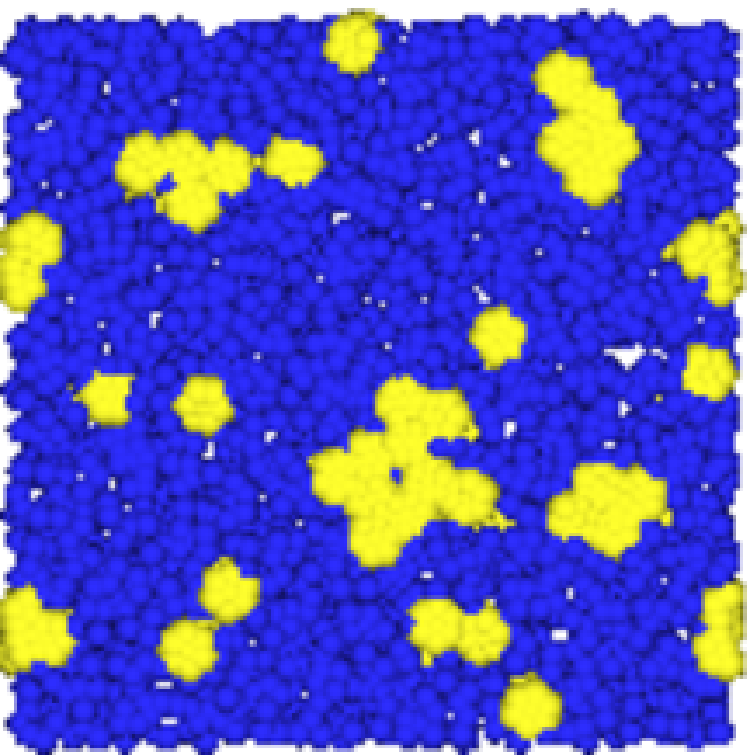}}
{\includegraphics[width=0.2\textwidth]{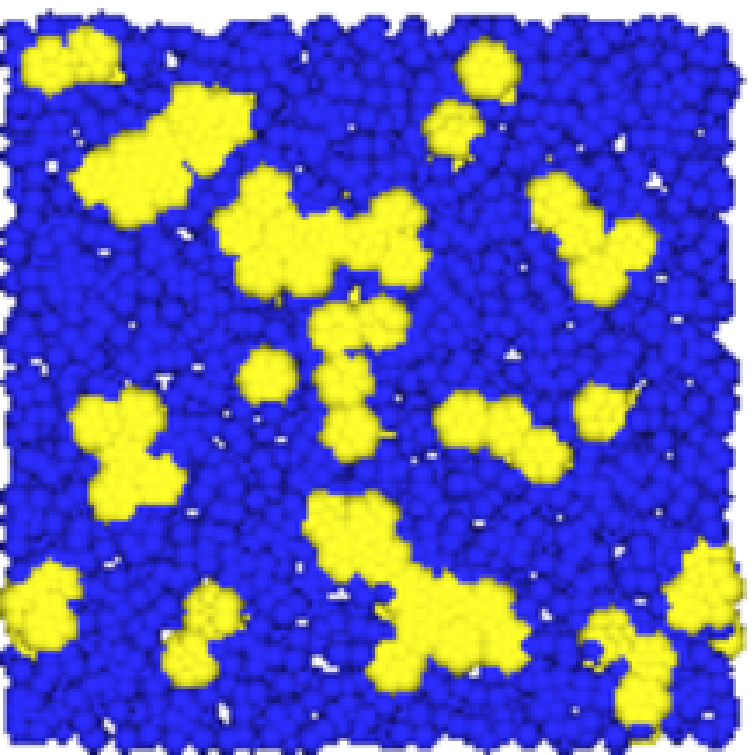}}
\caption{Top views of our flat bilayers $B_3$ (left panel) and $B_4$ (right panel) 
with transmembrane proteins. The bilayers $B_3$ and $B_4$ have the same number of 
proteins and areas of the vesicles $V_3$ and $V_4$, respectively.}
\label{fig:fig3}
\end{center}
\end{figure}

We have constructed three vesicles with initial bilayers of approximately 
equal surface areas and different protein concentrations. 
We name these vesicles $V_2$, $V_3$ and $V_4$, which have been formed 
inside $N_{w}\approx 161000$ water molecules. The numbers of lipid and 
protein molecules in our models have been given in Table \ref{tab:table1}. 
Fig. \ref{fig:fig2} demonstrates three dimensional views of these vesicles 
and their cross sections with the largest diameter so that the majority 
of clusters are visualized. We have also shown some water particles 
inside and outside vesicles. It is seen that larger protein clusters have 
induced lower curvatures in their neighborhood, and consequently, prominent 
deformations in their host vesicles. 

\begin{table}
\caption{\label{tab:table1}The protein and lipid content, and averaged 
curvatures of the simulated vesicles $V_{1}$--$V_{4}$. The quantity 
$\bar p_b$ has been computed for the bilayers $B_2$--$B_4$.}
\begin{ruledtabular}
\begin{tabular}{ccccc}
         & $V_1$ & $V_2$ & $V_3$ & $V_4$ \\
\hline
$N_l$ & 2300 & 2120  &  1940  & 1760 \\
$N_p$ & 0     & 20    & 40    &  60   \\
$\bar H_p$ & - & 0.0460 & 0.0381 & 0.0418 \\
$\bar H_l$ & 0.0564 & 0.0549 & 0.0582 & 0.0567 \\
$\bar H$ & 0.0564 & 0.0541 & 0.0544 & 0.0527 \\
$\tilde H$ & 0.0068 & 0.0135 & 0.0155 & 0.0158 \\
$\bar K$ & 0.0030 & 0.0029 & 0.0029 & 0.0027 \\
$\tilde K$ & 0.0014 & 0.0015 & 0.0015 & 0.0015 \\
$\bar p$ & - & 1.66 & 5 & 4.61 \\ 
$\bar p_b$ & - & 1.53 & 2.66 & 3.52 \\
\hline
$Q_c$  & 0 &  3  &   5  &   8  \\
$Q_f$  & 0 &  9  &   3  &   5  \\
$p(1)$ & - &  6  &  14  &  13  \\
$p(2)$ & - &  3  &  11  &  11  \\
$p(3)$ & - &  2  &   5  &  11  \\
$p(4)$ & - &  -  &   5  &   6  \\
$p(5)$ & - &  -  &   2  &   6  \\
$p(6)$ & - &  -  &   -  &   4  \\
$p(7)$ & - &  -  &   -  &   2  \\
$p(8)$ & - &  -  &   -  &   2  \\
\end{tabular}
\end{ruledtabular}
\end{table}
\begin{figure*}
\begin{center}
{\includegraphics[width=0.45\textwidth]{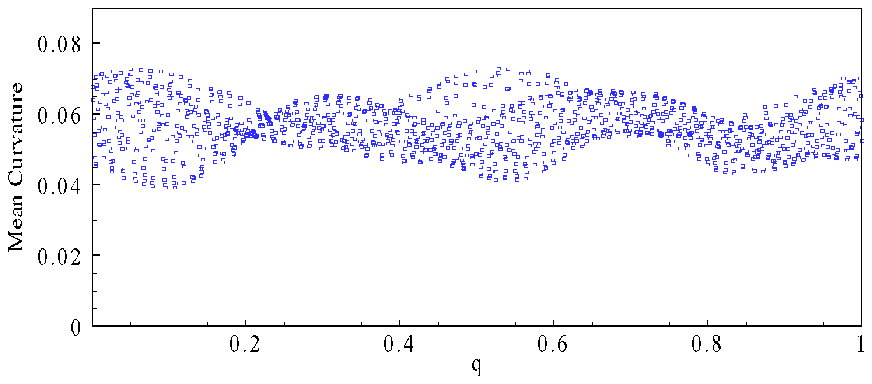}}
{\includegraphics[width=0.45\textwidth]{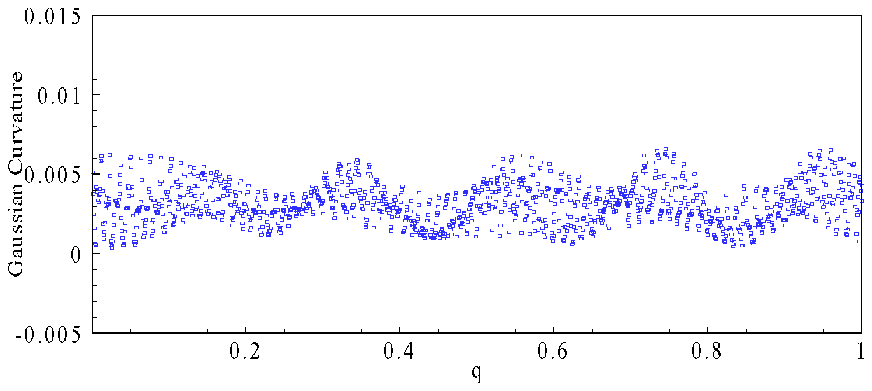}} \\
\vspace{0.1in}
{\includegraphics[width=0.45\textwidth]{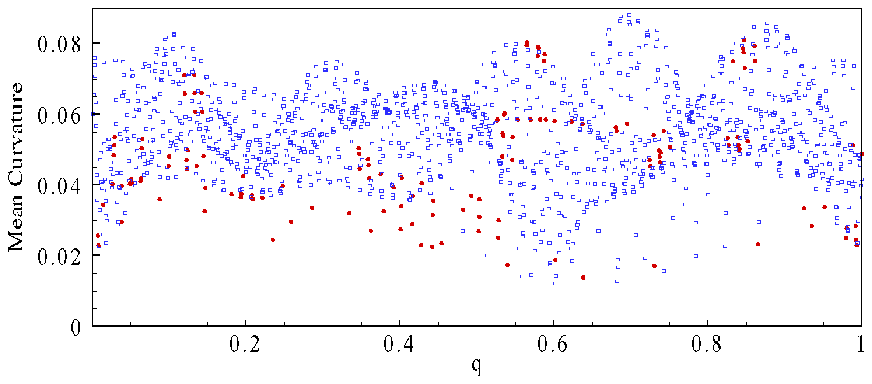}}
{\includegraphics[width=0.45\textwidth]{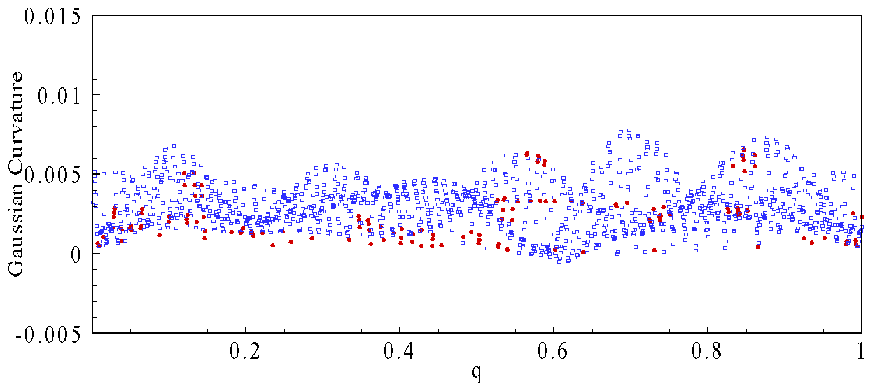}} \\
\vspace{0.1in}
{\includegraphics[width=0.45\textwidth]{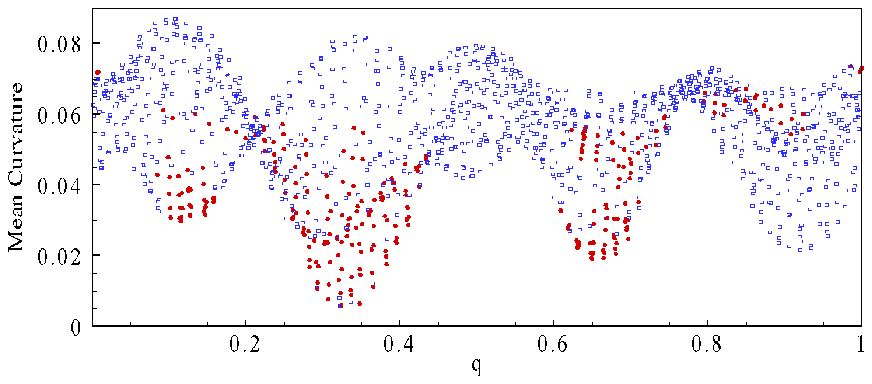}}
{\includegraphics[width=0.45\textwidth]{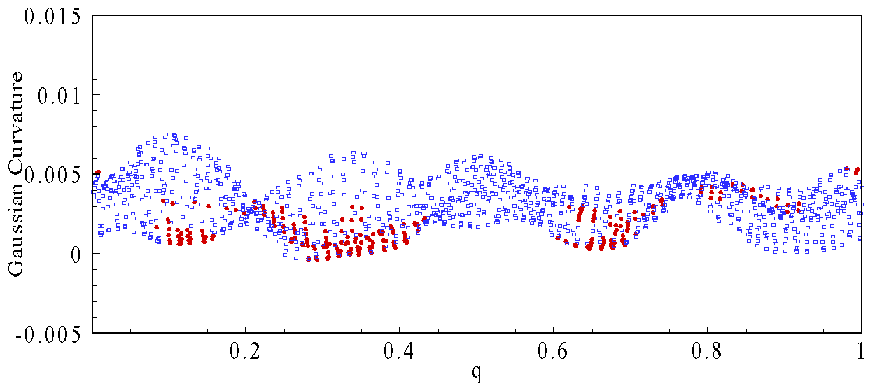}} \\
\vspace{0.1in}
{\includegraphics[width=0.45\textwidth]{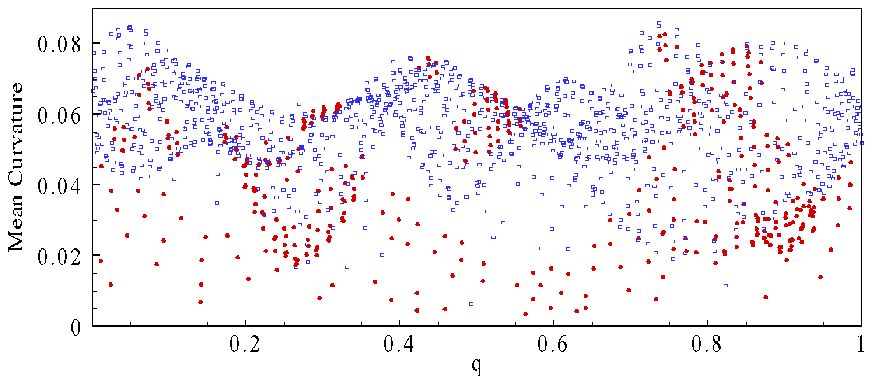}}
{\includegraphics[width=0.45\textwidth]{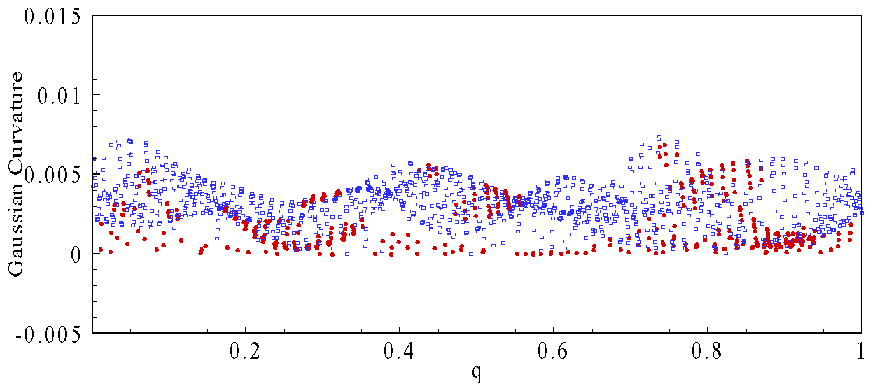}}
\caption{Scattered circles and squares mark, respectively, 
the mean curvature $H$ (left panels) and the Gaussian curvature 
$K$ (right panels) at the location of proteins (filled red circles) 
and lipid heads (blue squares). The averaged curvatures 
$\bar H_{p}$ and $\bar H_{l}$ as well as the average Gaussian 
curvature $\bar K$ and its standard deviation $\tilde{K}$ have 
been given in Table \ref{tab:table1}. From top to bottom: models 
$V_1$, $V_2$, $V_3$ and $V_4$. The horizontal axis variable is 
the normalized real number $q=n/N$ for the $n$th surface particle. 
}
\label{fig:fig4}
\end{center}
\end{figure*}

We also investigate several protein-embedded flat bilayers
where the aggregation of proteins is caused mainly by the 
depletion force. Comparing the sizes and population of clusters 
formed in bilayers and vesicles helps us to better understand 
the roles of entropy- and curvature-driven aggregation of proteins 
during vesiculation. We simulate three flat bilayers, which extend
to the sides of the simulation box. The surface areas of these bilayers 
and the population of their proteins match those of the vesicles 
$V_2$--$V_4$. We label these bilayers as $B_2$, $B_3$ and $B_4$. 
They remain in a flat equilibrium state because of periodic boundary 
conditions that keep them in touch with the sides of the simulation 
box. Fig. \ref{fig:fig3} displays the snapshots of $B_3$ and $B_4$.

We split the molecules of the outer layer of vesicles into two 
groups of lipids and proteins, and denote by $H_l(n)$ and $H_p(n)$ 
the mean curvatures at the locations of lipid and protein heads, 
respectively. We track only initially tagged particles living on 
the outer surface of vesicle (because the probability of flip-flop 
motions is low) and compute the local curvature having their 
coordinates $(r_n,\phi_n,\theta_n)$ for $n=1,2,\ldots,N$. 
We have plotted $H_{\mu}$ and $K_{\mu}$ ($\mu\equiv l,p$) in 
Fig. \ref{fig:fig4} for models $V_1$--$V_4$. The curvatures at 
different surface points have been marked by different symbols 
and colors. The scattered plots provide a quantitative insight 
to the effect of proteins on the vesicle shape. It is seen 
that the fluctuations of $H$ are higher in models with proteins. 
As we mentioned before, there is no correlation between the location 
of particles and their corresponding identifier $n$. Therefore, 
in Fig. \ref{fig:fig4}, points with close values of $n$ are not 
close physically. The oscillatory behavior of the graphs could 
change by renumbering the surface points but the major minima 
always coincide with protein clusters, which flatten their host 
vesicles locally.

We define $\bar H_l$ and $\bar H_p$ as the averages of mean 
curvatures (taken over the particles of the same kind) corresponding 
to lipid and protein heads, respectively. The average mean curvature 
$\bar H$ for all surface particles (protein and lipid heads) and its 
corresponding standard deviation $\tilde H$ have been given in 
Table \ref{tab:table1} together with $\bar H_l$ and $\bar H_p$. 
Lower values of $\bar H$ for models $V_{2}$--$V_{4}$ confirm that 
transmembrane proteins reduce the averaged mean curvature through 
creating low-curvature clusters. During vesiculation, protein 
molecules aggregate and form clusters. The number of clusters and 
the population of proteins in each cluster are determined by 
(i) the random arrangement of proteins in the generating bilayer 
membrane (ii) the interplay between depletion force and curvature 
induced interactions (iii) system temperature and the concentration 
of protein molecules. If protein molecules are scattered in a vesicle, 
the membrane will maintain its sphericity. However, if the same number 
of proteins build a single cluster, the highest variation in vesicle 
shape will be observed. In our simulations we have not seen these 
two extreme cases. Several clusters with different populations of 
proteins are usually formed in our vesicles.

\subsection{Curvature-mediated clustering}
\label{subsec:curvature-mediated-clustering}

Let us denote the number of scattered free proteins of a model by 
$Q_f$, and the number of its clusters by $Q_c$. Moreover, 
we indicate by $p(m)$ the number of proteins in the $m$th cluster. 
For models $V_2$, $V_3$ and $V_4$, we have reported 
the values of $Q_f$, $Q_c$ and $p(m)$ in Table \ref{tab:table1}. 
It is seen that the biggest cluster have been formed in vesicle $V_{3}$ and 
not in $V_{4}$, which has more proteins. Consequently, $\bar H_p$ is lower 
in vesicle $V_{3}$ than $V_{4}$. The standard deviation $\tilde{H}$ is 
$\approx 10\%$ of $\bar H$ for the near-spherical vesicle $V_{1}$, but it 
increases remarkably for protein-embedded vesicles. As clusters grow, the 
curvature associated with the ensemble of lipids increases, and the 
vesicle becomes more aspherical. This can be understood from the larger 
value of $\bar H_l$ in vesicle $V_{3}$. A reverse phenomenon is also 
possible: if during the vesicle formation the curvature decreases in 
certain regions, proteins will migrate there and form low-curvature 
clusters. To show the correlation between the number of proteins in 
clusters and $\bar H$, we have computed the quantity 
\begin{eqnarray}
\bar p = \frac {1}{Q_f+Q_c} \left [ Q_f + \sum_{m=1}^{Q_c} p(m) \right ],
\label{eq:define-bar-p}
\end{eqnarray}
and given its magnitude in Table \ref{tab:table1}. We also define $\bar p_b$ 
using a formula similar to (\ref{eq:define-bar-p}), but for our model bilayers 
$B_i$ that correspond to vesicles $V_i$ ($i=2,3,4$). The computed values of 
$\bar p_b$ are given in Table \ref{tab:table1}. 

Since bilayers remain flat during simulation, the effect of membrane curvature 
on the aggregation process is ignorable and $\bar p_b$ indicates the contribution 
of the depletion force to cluster formation. Comparing the values 
of $\bar p$ and $\bar p_b$ clearly shows that curvature induced interactions, 
during vesiculation, facilitate the cluster growth and we get $\bar p > \bar p_b$. 
Moreover, $\bar p_b$ is a monotonic function of $N_p$ while $\bar p$ is not. 
The reason is the lack of an effective fragmentation mechanism in flat bilayers: 
larger protein concentrations always lead to bigger clusters. In vesicles, 
however, the tendency to form a structure with minimum bending energy leads to 
the fragmentation of big clusters at the turning (maximum) point of the 
function $\bar p(N_p)$.

\begin{table}
\caption{\label{tab:table2}The number and sizes of protein clusters 
for vesicles $V_{4}$--$V_{6}$ with $N_p=60$ and different initial 
random arrangements of proteins.}
\begin{ruledtabular}
\begin{tabular}{cccc}
         & $V_4$ & $V_5$ & $V_6$ \\
\hline
$\bar p$ & 4.61 & 4.61 & 4.61 \\ 
$Q_c$  & 8 & 9 & 9  \\
$Q_f$  & 5 & 4 & 4  \\
$p(1)$ & 13 & 14 & 14  \\
$p(2)$ & 11 & 11 & 9  \\
$p(3)$ & 11 & 8 & 8  \\
$p(4)$ & 6 & 7 & 7  \\
$p(5)$ & 6 & 6 & 6  \\
$p(6)$ & 4 & 3 & 5  \\
$p(7)$ & 2 & 3 & 3  \\
$p(8)$ & 2 & 2 & 2  \\
$p(9)$ & - & 2 & 2  \\
\end{tabular}
\end{ruledtabular}
\end{table}

For all surface points including lipid and protein heads, we have also 
calculated (see Table \ref{tab:table1}) the average Gaussian curvature 
$\bar K$ and its standard deviation $\tilde{K}$ using (\ref{eq:K-and-H})
and (\ref{eq:averaged-curvatures}). Right panels of Fig. \ref{fig:fig4} 
show the distribution of $K$ for lipid and protein heads. Based on the 
Gauss-Bonnet theorem, any compact manifold ${\cal M}$ without boundary, 
is topologically equivalent to a sphere and the surface integral 
$\int_{{\cal M}} K ~{\rm d}A$ of Gaussian curvature will be invariant. 
Since our vesicles become aspherical through the shape transformations 
induced by protein clustering, Gauss-Bonnet theorem applies and all models 
$V_1$--$V_4$ must be topologically equivalent. Given that the head groups 
of proteins and lipids are identical in our simulations and the areas of 
generating bilayers are almost the same, the surface integral will be 
approximately equal to $A \bar K$. Data in Table \ref{tab:table1} show 
that both $\bar K$ and $\tilde K$ remain invariant (from one vesicle to 
another) within a reasonable error threshold. This confirms the 
self-consistency of our models and the results obtained from spherical 
harmonic expansions.

\subsection{Convergence tests}
\label{subsec:convergence}

We have continued our simulations until the size and number of clusters 
become constant in a relaxed equilibrium state. To assure that simulated 
vesicles are in equilibrium, we have carried out various experiments. 
In the first experiment we generated two vesicles from the same progenitor 
bilayer of $V_4$, but using two different sets of randomly distributed 
proteins. We have given the properties of new vesicles $V_5$ and $V_6$
in Table \ref{tab:table2}. Although the vesicles $V_4$, $V_5$ and $V_6$ 
start from different initial conditions, there are minor differences 
between the properties of their clusters. Notably, they posses the same 
shape indicator $\bar p$=$4.61$. 

\begin{figure}
\begin{center}
{\includegraphics[width=0.2\textwidth]{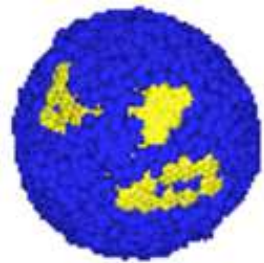}}
\caption{Three dimensional views of dissociated clusters in vesicle 
$V_7$ with $N_p=20$ protein molecules initially located in a single big cluster.}
\label{fig:fig5}
\end{center}
\end{figure}

To demonstrate that we usually reach
a physical equilibrium and not a kinetically trapped state, we designed 
a second experiment and produced a vesicle from an initial bilayer where 
$N_p$=$20$ proteins (similar to vesicle $V_2$) had formed an initial big 
cluster. In the resulting vesicle $V_7$, the proteins of the initial single 
cluster are dissociated into three smaller separate clusters as shown 
in Fig. \ref{fig:fig5}. This result is, again, consistent with the general 
features of $V_2$, which had been obtained from a completely different 
initial condition. It is worth noting that we have observed a transient 
interplay between clustering and fragmentation well before reaching the 
equilibrium state.

We have repeated our simulations with $\rho=0.8$ and with non-flexible 
lipid and protein chains. Using a bending stiffness $K_{\rm bend}=5$ 
in lipids and $K_{\rm bend}=80$ in protein strands, the bilayer 
$\rightarrow$ vesicle transition is slowed down but we do not 
observe considerable change, either quantitative or qualitative, 
in the clustering phenomenon and the shape transformation of vesicles: 
the numbers and sizes of final clusters and the shape parameter 
$\bar p$ are similar in all models. By making stiffer molecular 
chains and increasing the density, lipid diffusion is decreased, 
which in turn, yields a longer relaxation time.


\section{Conclusions}
\label{sec:conclusions}

We have studied the phenomena of protein clustering and membrane shape 
transformation during membrane vesiculation and afterwards. Comparing 
relaxed vesicles and bilayers shows that protein clustering during 
vesiculation occurs due to both entropy-driven depletion force and 
the curvature-mediated interactions. The latter effect enhances the 
generation of larger protein clusters and determines bilayer's bending 
rigidity. Once the vesicle is formed, protein clusters locally flatten 
their host vesicles and increase the bending energy as $\bar H$ and 
$\bar H_l$ increase.  The system, however, cannot tolerate the increase 
in the bending energy for protein concentrations beyond a critical value. 
By increasing the protein concentration, bigger protein clusters are 
not formed in our simulations, or they break apart. We anticipate a 
uniform distribution of fragmented clusters, like the shape of a 
soccer ball. Our observations show that low protein concentrations 
do not lead to efficient cluster formation.

\begin{acknowledgments}
This work was partially supported by the research vice-presidency
at Sharif University of Technology. We thank the referees for their
constructive comments, which substantially improved the paper.  
\end{acknowledgments}

\end{document}